\documentclass[twocolumn,floatfix,preprintnumbers,nofootinbib,superscriptaddress]{revtex4}

\usepackage{ulem}
\usepackage{bm}
\usepackage{times}
\usepackage{amssymb,amsbsy,amsmath,amsfonts}
\usepackage{graphicx}
\usepackage{float}
\usepackage{color}
\usepackage{morefloats}
\usepackage{rotating}
\usepackage{srcltx}
\usepackage{slashed}
\usepackage{subfigure}
\usepackage{multirow}
\usepackage{verbatim}
\usepackage{hyperref}
\usepackage{tabularx}

\usepackage[outdir=./]{epstopdf}

\begin{document}

\title{Electromagnetic form factors of neutron and neutral stange hyperons in the oscillating view of point}

\author{An-Xin Dai}
\affiliation{Institute of Modern Physics, Chinese Academy of Sciences, Lanzhou 730000, China}
\affiliation{School of Nuclear Science and Technology, University of Chinese Academy of Sciences, Beijing 101408, China}
\author{Zhong-Yi Li}
\affiliation{Institute of Modern Physics, Chinese Academy of Sciences, Lanzhou 730000, China}
\affiliation{School of Nuclear Science and Technology, University of Chinese Academy of Sciences, Beijing 101408, China}
\author{Lei Chang}~\email{leichang@nankai.edu.cn}
\affiliation{School of Physics, Nankai University, Tianjin 300071, China}
\author{Ju-Jun Xie}~\email{xiejujun@impcas.ac.cn}
\affiliation{Institute of Modern Physics, Chinese Academy of Sciences, Lanzhou 730000, China}
\affiliation{School of Nuclear Science and Technology, University of Chinese Academy of Sciences, Beijing 101408, China}
\affiliation{School of Physics and Microelectronics, Zhengzhou University, Zhengzhou, Henan 450001, China}

\date{\today}

\begin{abstract}

Based on the recently precise measurements of the electron-positron
annihilation reactions into a neutron and anti-neutron pair by
BESIII collaboration, the effective form factor of neutron was
determined in the time like region, and it was found that the
effective form factor of neutron is smaller than the ones of proton.
The effective form factors of neutron shows a periodic behaviour,
similar to the case of proton. Here, a compared analysis for
$\Lambda$, $\Sigma^0$ and $\Xi^0$ is performed. Fits of the available data on
the effective form factors of $\Lambda$, $\Sigma^0$ and $\Xi^0$ with charge
zero, allow to show a universal phenomenon of the oscillation behavior in their effective form factors. However, this needs to be confirmed by
future precise experiments. Both theoretical and experimental
investigations of this phenomenon can shed light on the reaction
mechanisms of the electron-positron annihilation process.

\end{abstract}
\maketitle

\section{Introduction}

The investigation of baryon structure is one of the most important
issues and is attracting much attention. Electric $G_E$ and magnetic
$G_M$ form factors (EMFFs) are fundamental quantities that describe
the electromagnetic structure of
hadrons~\cite{Brodsky:1974vy,Geng:2008mf,Green:2014xba,Pacetti:2014jai}.
A lot of the experimental and theoretical efforts have been made in
the past decades. On the one hand, with the planed and upgrade of
experimental facilities, the EMFFs of proton at the space like
region can be extracted from the $ep$
scattering~\cite{Anderle:2021wcy}. On the other hand, the
measurements of time-like region EMFFs of hadrons can be done in the
electron-position annihilation process, which provides a key to
understanding quantum chromodynamics effects in bound states. For
example, there are great progress in the study of baryon EMFFs in
the time like regions, both on the
experimental~\cite{Akhmetshin:2015ifg,Andreotti:2003bt,Antonelli:1998fv,Ablikim:2019eau,Bardin:1994am,Bisello:1983at,Ambrogiani:1999bh,Aubert:2005cb,Lees:2013uta,Lees:2013ebn,Ablikim:2015vga,BESIII:2020uqk,BaBar:2007fsu,BESIII:2017hyw,BESIII:2019nep,BaBar:2013ves,BESIII:2019hdp,BESIII:2021rqk,BESIII:2021ccp,BESIII:2021dfy,BESIII:2021cvv,BESIII:2021aer}
and theoretical
sides~\cite{Gousset:1994yh,Yang:2019mzq,Ramalho:2019koj,Yang:2020rpi,Haidenbauer:2020wyp,Lin:2008mr,Liu:2009mb,Kubis:2000aa,Iachello:1972nu,Iachello:2004aq,Bijker:2004yu,Baldini:2007qg,Faldt:2016qee,Haidenbauer:2016won,Faldt:2017kgy,Yang:2017hao,Denig:2012by,Bianconi:2015owa,Tomasi-Gustafsson:2020vae,Bystritskiy:2021frx,Li:2021lvs,Lin:2021xrc,Cao:2021asd}.

Very recently, the reaction of $e^+e^- \to n\bar{n}$ was measured by
the BESIII experiment at centre of mass energies between 2.00 and
3.08 GeV, with very high precision. This precise measurement
clarifies that the effective form factors of neutron are of the same
magnitude but smaller than the ones of the proton. This new result
shows that the photon-proton interaction is stronger than the
corresponding photon-neutron interaction. On the other hand, an
oscillating behavior in the modulus of the effective form factors
after the subtraction of a dipole function has been observed for the
neutron~\cite{BESIII:2021tbq}, similarly to the case of
proton~\cite{BaBar:2013ves,BESIII:2019hdp,BESIII:2021rqk,Bianconi:2015owa,Tomasi-Gustafsson:2020vae,Bystritskiy:2021frx}.
Further studies are needed to explain this oscillation behaviors of
the effective form factors of the nucleon.

\begin{figure}[htbp]
	\centering \vspace{-0.5cm}
	\includegraphics[scale=0.45]{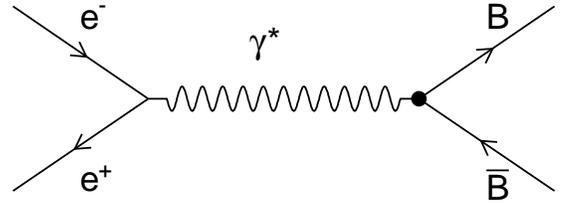}
	\vspace{-1.5cm}
	\caption{The feynman diagram for the reaction of $e^+ e^- \to B\bar{B}$.}
	\label{fig:fdg}
\end{figure}

Assuming one-photon exchange (see fig.~\ref{fig:fdg}), the so-called
Born cross section of electron-positron annihilation into a
baryon-anti-baryon pair, $e^+ e^- \to B\bar{B}$ with $B$ a spin
$1/2$ baryon and $\bar{B}$ a anti-baryon, can be expressed in terms
of the time like electric and magnetic form factors $G_{E}$ and
$G_{M}$ as~\cite{Denig:2012by}~\footnote{More general, the original
formula has been written in terms of the Dirac and Pauli form
factors, $F_1$ and $F_2$, with $G_E = F_1 + \tau F_2$ and $G_M = F_1
+ F_2$.}
\begin{equation}
\sigma_{B} (s) = \frac{4 \pi \alpha^2
\beta}{3 s^2} \left( s \left|G_{M}\left( s \right)\right|^{2} +
2M^2_B \left|G_{E}\left( s \right)\right|^{2}\right),
    \label{eq:tcs}
\end{equation}
where $\alpha = e^2/(4\pi) = 1/137.036$ is the fine-structure
constant and $\beta = \sqrt{1-4 M_B^{2}/s}$ is a phase-space factor,
$M_B$ is the baryon mass. While $s$ is the invariant mass square of
the $e^+ e^-$ system. In the space like region, $G_E$ and $G_M$ are
real, while in the time like region, they are complex.

The measurement of the total cross section in Eq.~\eqref{eq:tcs} at
a fixed energy allows for determination of the combination of
$|G_E|^2$ and $|G_M|^2$. Instead of a separation between $G_E$ and
$G_M$, one can easily obtain the effective form factor $G_{\rm
eff}(s)$ from the total cross section of $e^+ e^-$ annihilation
process~\cite{BESIII:2017hyw,BESIII:2019nep}. It is defined as
\begin{eqnarray}
|G_{\mathrm{eff}}\left( s \right)|  &=&  \sqrt{\frac{\sigma_B}{[1+1/(2\tau)][4\pi\alpha^2\beta/(3s)]}} \nonumber \\
& = & \sqrt{\frac{2
\tau\left|G_{M}\left(q^{2}\right)\right|^{2}+\left|G_{E}\left(q^{2}\right)\right|^{2}}{1+2
\tau}},
\end{eqnarray}
where $\tau = s/(4 M^2_B)$. The effective form factor $|G_{\rm
eff}(s)|^2$ is a linear combination of $|G_E|^2$ and $|G_M|^2$, and it indicates how much the experimental $e^+e^- \to B \bar{B}$ cross section differs from a point-like baryon $B$. Therefore, the effective form factor $G_{\rm eff}(s)$ can be determined from a measurement of the total cross section of the $e^+ e^-$ annihilation process. The $|G_{\rm eff}(s)|$ values depend, in principle, on the kinematics where the measurement was performed. It describes also the coupling constants for the elementary process
$e^+e^-$ to quark-anti-quark pairs.

In general, the experimental data on the effective form factor
$|G_{{\rm eff}}(s)|$ can be well reproduced by a dipole function
$G_D(s)$, which has a decreasing behavior as a function of $s$. Its
form are commonly written
as~\cite{Bianconi:2015owa,Tomasi-Gustafsson:2020vae}
\begin{equation}
G_D(s)=\frac{c_0} {(1-\gamma s)^2 }, \label{eq:GD}
\end{equation}
where $c_0$ and $\gamma$ are constant, which will be determined by the experimental data.

On the other hand, for the oscillating part, it can be easily fitted by the following equation:
\begin{equation}
G_{\rm {osc }}(s) = \frac{c_0} {(1-\gamma s)^2 } A  \cos (C \sqrt{s}
+ D), \label{eq:Gosc}
\end{equation}
where $A$, $B$, $C$, and $D$ are constant and will be determined by
the experimental measurements. It is worth to mention that we take
the oscillating part as a function with the invariant mass
$\sqrt{s}$ rather than the relative momentum $p$ as used in
Refs.~\cite{Bianconi:2015owa,Tomasi-Gustafsson:2020vae,BESIII:2021tbq,BESIII:2021rkn},
where $p$ is a function of the invariant mass $\sqrt{s}$: $p =
\sqrt{s(\tau -1)}$. Furthermore, the natural exponential reduced
function used in
Refs.~\cite{Bianconi:2015owa,Tomasi-Gustafsson:2020vae,BESIII:2021tbq,BESIII:2021rkn}
is also replaced by the dipole function which is the same as for the
main part of the effective form factor.

Then, the total effective form factor $G_{\rm eff}$ is:
\begin{eqnarray}
G_{\rm {eff}}(s) &=& G_D(s) + G_{\rm {osc }}(s) \nonumber \\
 &=& \frac{c_0} {(1-\gamma s)^2 } \left( 1 + A  \cos (C \sqrt{s} + D) \right ) .
\end{eqnarray}
After subtracting the well-established dipole part of $G_D(s)$, one
can easily obtain the oscillation part of $G_{\rm {osc}}(s)$.

Although their is no fundamental explanation for the oscillation
behavior of nucleon effective form factors, modelling the joint
information from all the current experimental data on the effective
form factors of baryons may bring to a new view of the reaction
mechanism for the formation of baryons. Along this line, in this
work, we performed a combined analysis of the effective form factors of these neutral particles with spin $1/2$: neutron, $\Lambda$,
$\Sigma^0$, and $\Xi^0$. It is found
that the effective form factors of these baryons show a universal phenomenon of an oscillation behavior, similar to the
observation of nucleon. Further experimental and theoretical
investigations of such behavior may open a new window for the
electromagnetic structure of $\Lambda$, $\Sigma^0$, and $\Xi^0$ hyperons.

\section{Analysis of the effective form factors of neutron, $\Lambda$, $\Sigma^0$, and $\Xi^0$}

\subsection{Global fit to the effective form factor with a dipole function}

Firstly, we perform two parameters ($c_0$ and $\gamma$) $\chi^2$
fits with the dominant dipole part $G_D(s)$ as in Eq.~\eqref{eq:GD}
to the experimental data on the effective form factors of neutron,
$\Lambda$, $\Sigma^0$, and $\Xi^0$. The fitted parameters are shown
in Table~\ref{tab:GDpara}.

\begin{table}[htbp]
    \caption{Fitted parameters for the effective form factors with a dipole function shown in Eq.~\eqref{eq:GD}. The unit of $\gamma$ is ${\rm GeV}^{-2}$.} \label{tab:GDpara}
    \begin{ruledtabular}
        \begin{tabular}{ccccc}
            \textrm{Parameter}&
             \textrm{$n$}&
            \textrm{$\Lambda$}&
            \textrm{$\Sigma^0$}&
            \textrm{$\Xi^0$}\\
            \colrule
            $\gamma$     & $1.41$ (fixed)   & $0.34 \pm 0.08$   & $0.26 \pm 0.01$      & $0.21 \pm 0.02 $\\
            $c_0$        & $3.48 \pm 0.06$  & $0.11 \pm 0.01$   & $0.033 \pm 0.007$     & $0.023 \pm 0.008 $\\
            $\chi^2$/dof & 4.3                    & 2.4               & 1.1                & $3.0$ \\
        \end{tabular}
    \end{ruledtabular}
\end{table}

For the neutron, we take the value of $\gamma$ as in
Refs.~\cite{BESIII:2021tbq,BESIII:2021rkn}. This value is also used
for the proton. If we make $c_0$ and $\gamma$ are free parameters
for the case of neutron, there is always another solution from the
$\chi^2$ fit, which are: $c_0 = 96.6$ and $\gamma = 6.6$ ${\rm
GeV}^{-2}$, and both of them have huge errors. It is also found that
$c_0$ and $\gamma$ are strongly correlated.

\begin{figure*}[htbp]
    \centering
    \includegraphics[scale=0.5]{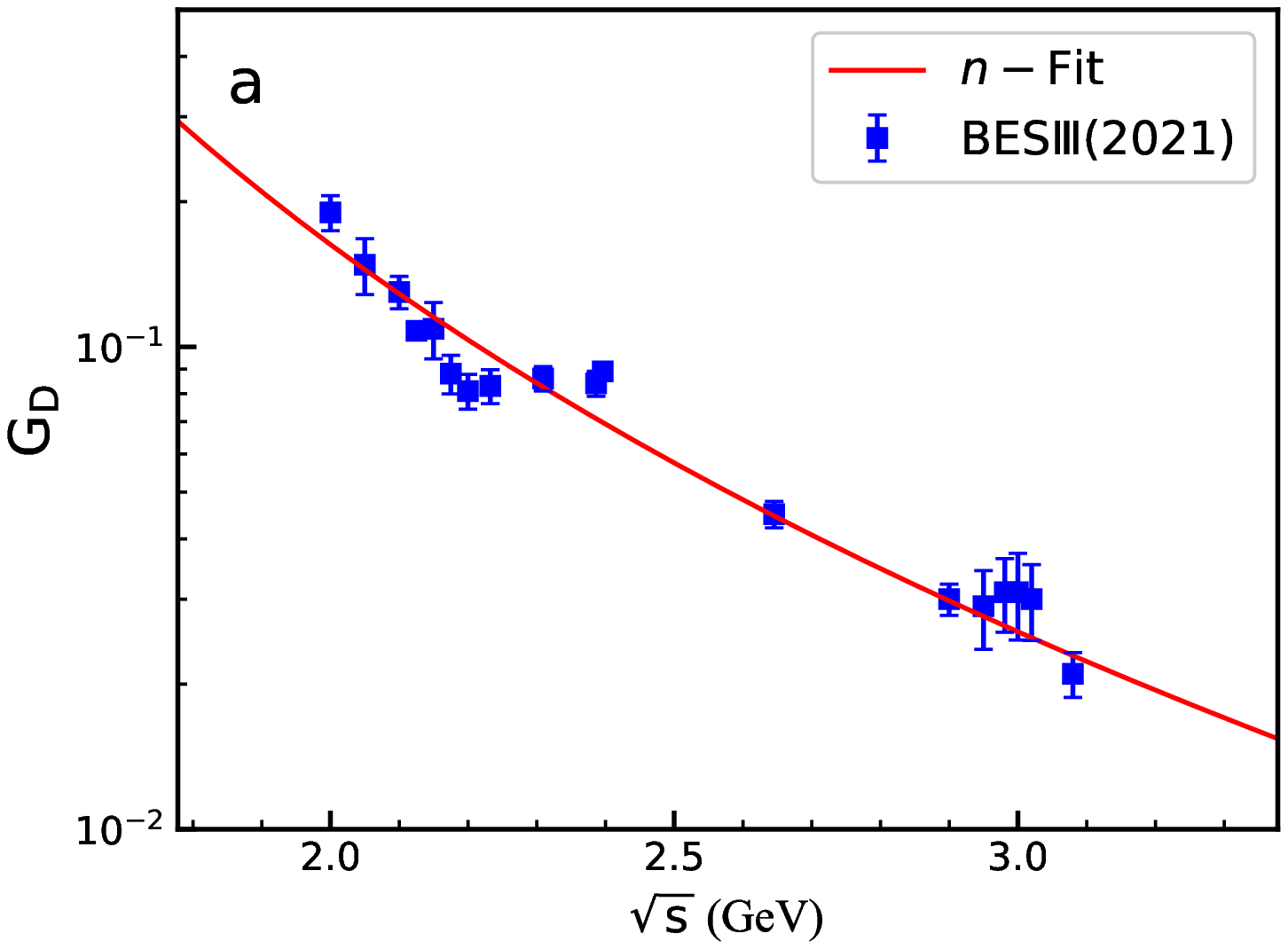}
    \includegraphics[scale=0.5]{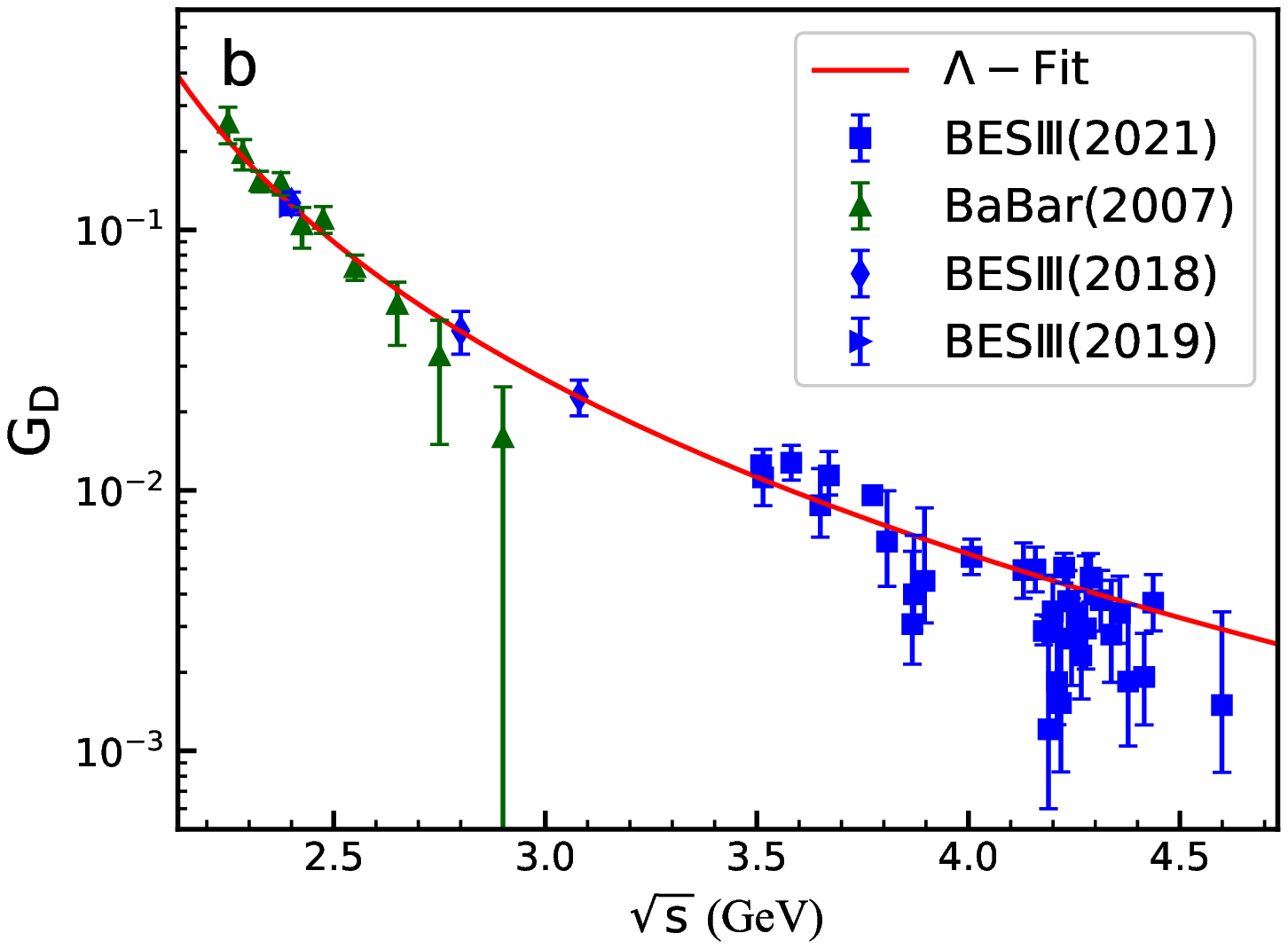}
    \includegraphics[scale=0.5]{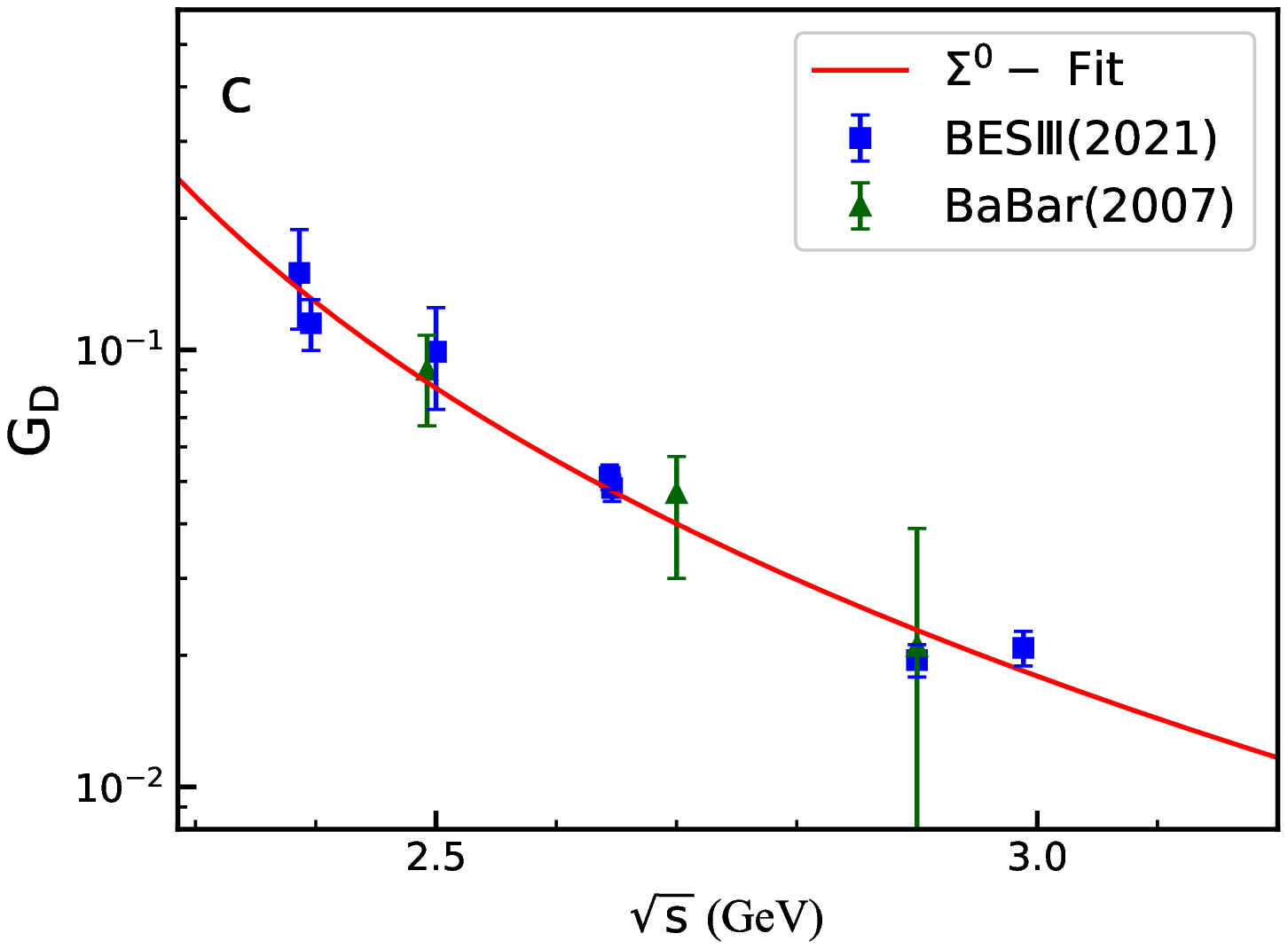}
    \includegraphics[scale=0.5]{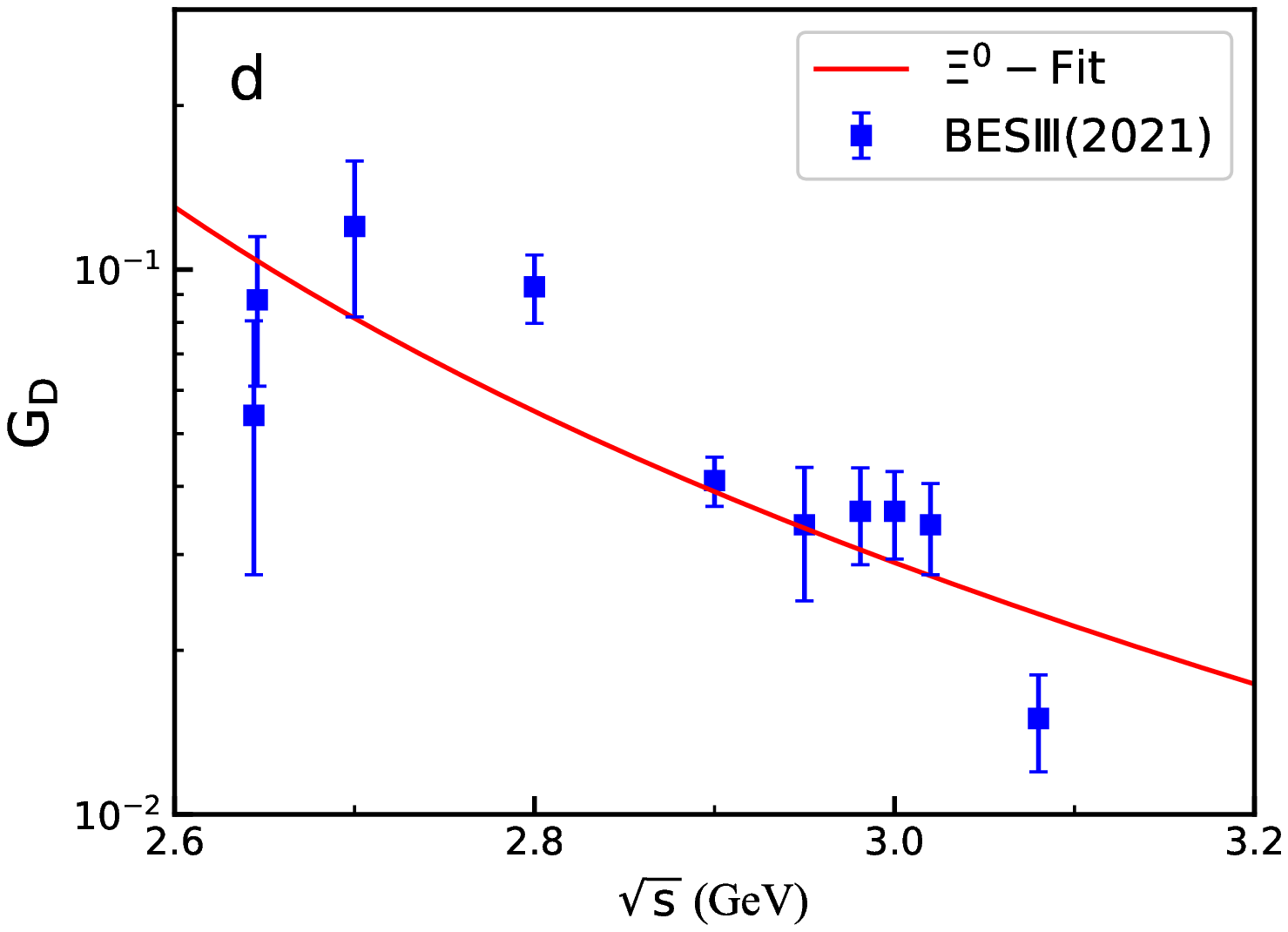}
    \caption{Fitted effective form factors of neutron (a), $\Lambda$ (b), $\Sigma^0$ (c) and $\Xi^0$ (d). The data are taken from BaBar collaboration~\cite{BaBar:2007fsu} and BESIII collaboration~\cite{BESIII:2017hyw,BESIII:2019nep,BESIII:2021ccp,BESIII:2021aer}.}
    \label{fig:LambdaGeff}
\end{figure*}

In Figs.~\ref{fig:LambdaGeff}, we show the fitted results for the
effective form factors of neutron, $\Lambda$, $\Sigma^0$, and
$\Xi^0$ with a dipole function $G_D(s)$ as in Eq.~\eqref{eq:GD},
respectively. One can see that the effective form factors can be
fairly well reproduced by using the dipole function.~\footnote{The
smooth behavior of the proton effective form factor is described
with a product of a free monopole and the standard dipole.}

One the other hand, from the fitted values of $\gamma$, it is found
that the effective form factors of $\Lambda$, $\Sigma^0$, and
$\Xi^0$ hyperons are reduced more fast than the one of nucleon, when
$\sqrt{s}$ is growing. For $\Lambda$, from the invariant mass
$\sqrt{s} = 2.3 $ GeV to $3.3$ GeV, $G_D(s)$ is reduced from $0.172$
to $0.015$, while for the case of neutron, from $\sqrt{s} = 2.0$ GeV
to $3.0$ GeV, its effective form factor is reduced from the value of
$0.162$ to $0.025$.

\subsection{Oscillations}

After subtracting the main dipole part from the effective form
factor, then one get the oscillation part $G_{\rm {osc }}(s)$. We
have performed three parameters ($A$, $C$, and $D$) $\chi^2$ fits
with the oscillation part $G_{\rm osc}(s)$ as in Eq.~\eqref{eq:Gosc}
for the neutron, $\Lambda$, and $\Sigma^0$. The fitted parameters
are shown in Table~\ref{tab:Goscpara}. Since the uncertainties of
these experimental data are large, the fitted parameters have big
uncertainties. In addition, the parameters $C$ and $D$ are strongly
correlated.

\begin{table}[htbp]
    \caption{Fitted parameters for the oscillation part of the effective form factors of $\Lambda$, $\Sigma^0$, and neutron. The unit of $C$ is ${\rm GeV}^{-1}$.}  \label{tab:Goscpara}
    \begin{ruledtabular}
    \begin{tabular}{cccc}
    \textrm{Parameter}&
    \textrm{neutron} &
    \textrm{$\Lambda$}&
    \textrm{$\Sigma^0$}\\
    \colrule
    $A$           & $0.176 \pm 0.027$   & $0.235 \pm 0.033$      & $0.199 \pm 0.102$      \\
    $C$           & $12.16 \pm 0.61$   & $3.99 \pm 0.14$        & $11.94 \pm 1.13$      \\
    $D$           & $1.66 \pm  1.41$   & $4.61 \pm 0.40$        & $0.99 \pm 3.13$      \\
    \end{tabular}
    \end{ruledtabular}
\end{table}

\begin{figure*}[htbp]
    \centering
    \includegraphics[scale=0.42]{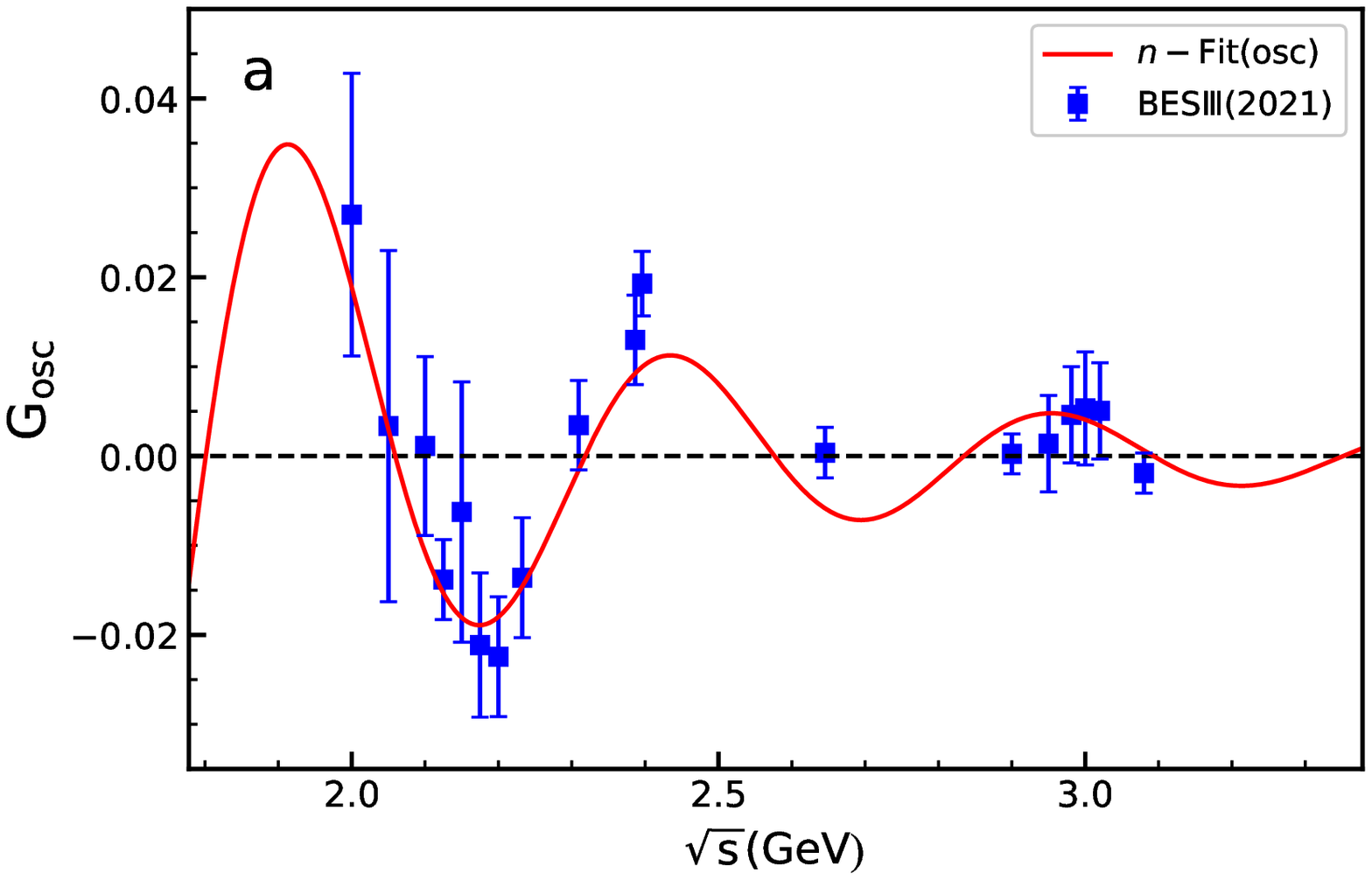}
    \includegraphics[scale=0.42]{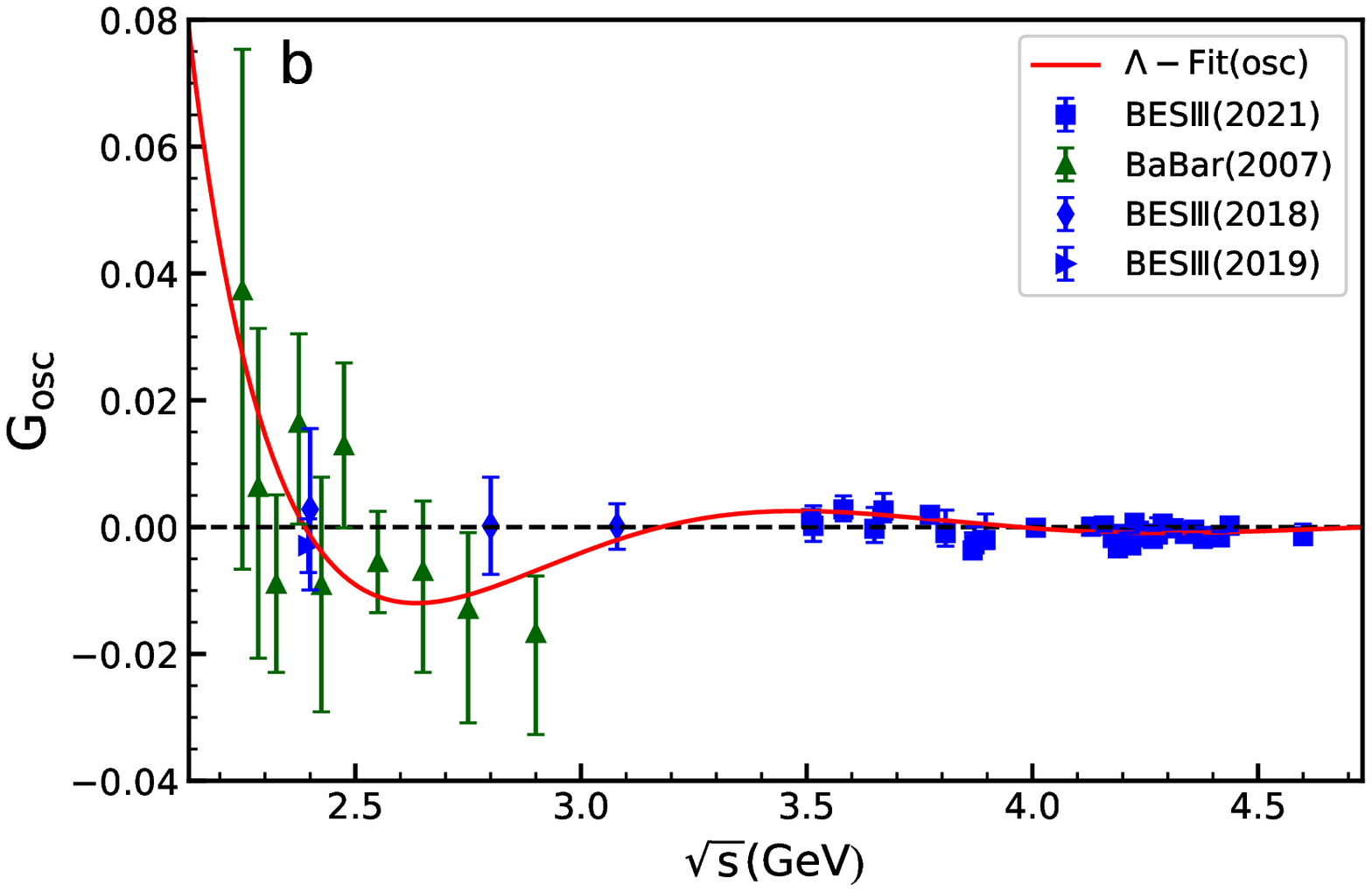}
    \includegraphics[scale=0.42]{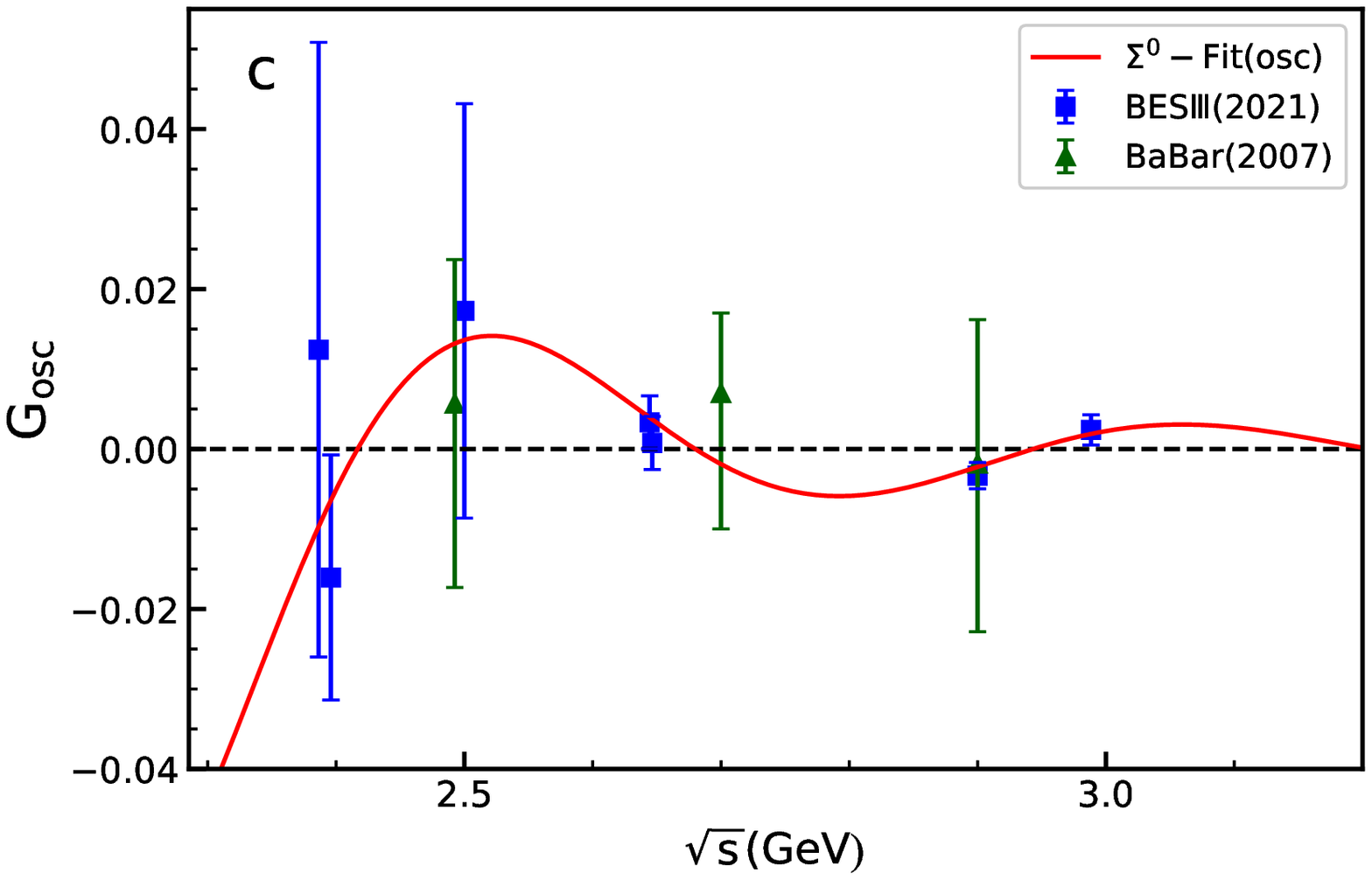}
    \includegraphics[scale=0.42]{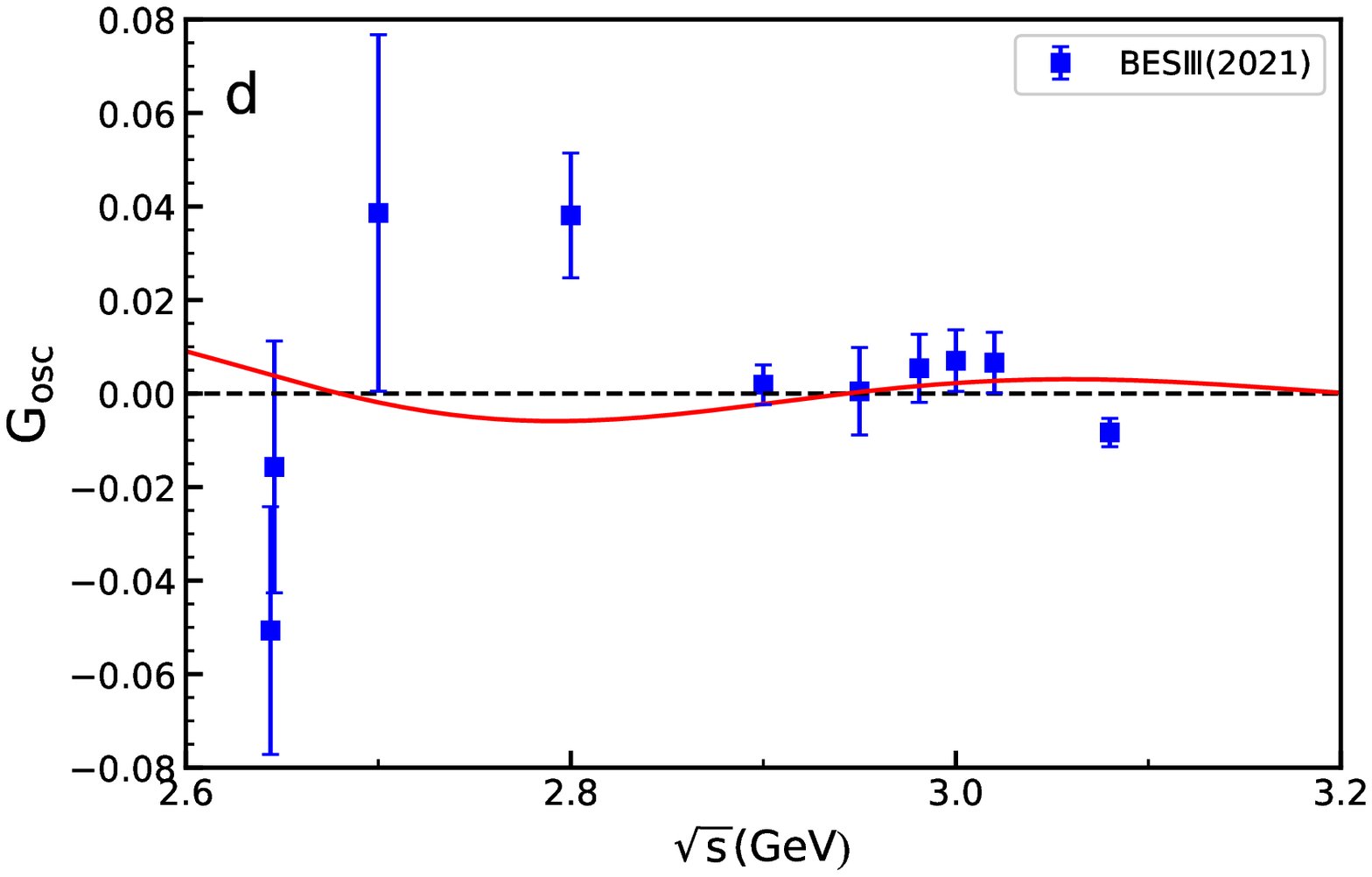}
    \caption{Fitting results of oscillation part of neutron (a), $\Lambda$ (b), $\Sigma^0$ (c), and $\Xi^0$ (d) effective form factors.}
    \label{fig:LambdaGosc}
\end{figure*}

With the central values of these fitted parameters shown in
Tab.~\ref{tab:Goscpara}, in Fig.~\ref{fig:LambdaGosc}, we show the
numerical results for the oscillation part of the effective form
factors of neutron, $\Lambda$, $\Sigma^0$, and $\Xi^0$. One can see
that the oscillation part can be fairly well described by a function
$G_{\rm osc}(s)$ as in Eq.~\eqref{eq:Gosc}. It is worth to mention
that we can not obtain a good fit to the oscillation part of the
$\Xi^0$ hyperon since the large errors of the experimental data. In
Fig.~\ref{fig:LambdaGosc} (d), the red line stands for the fitted
results of the $\Sigma^0$.

From Figs.~\ref{fig:LambdaGosc} (b), (c), and (d), it is found that
the data for the $\Lambda$, $\Sigma^0$, and $\Xi^0$ have larger
errors in the threshold region. In our fit for the case of
$\Sigma^0$, the first date was not considered. It is expected that
the future more precise experimental measurements will check the
oscillation behavior of the effective form factor of the $\Lambda$,
$\Sigma^0$, and $\Xi^0$ hyperons. Moreover, the process of
hyperon-antihyperon pairs production in the electron-position
annihilation can be also used to study the polarizations of
hyperons.

\begin{figure*}[htbp]
    \centering
    \includegraphics[scale=0.7]{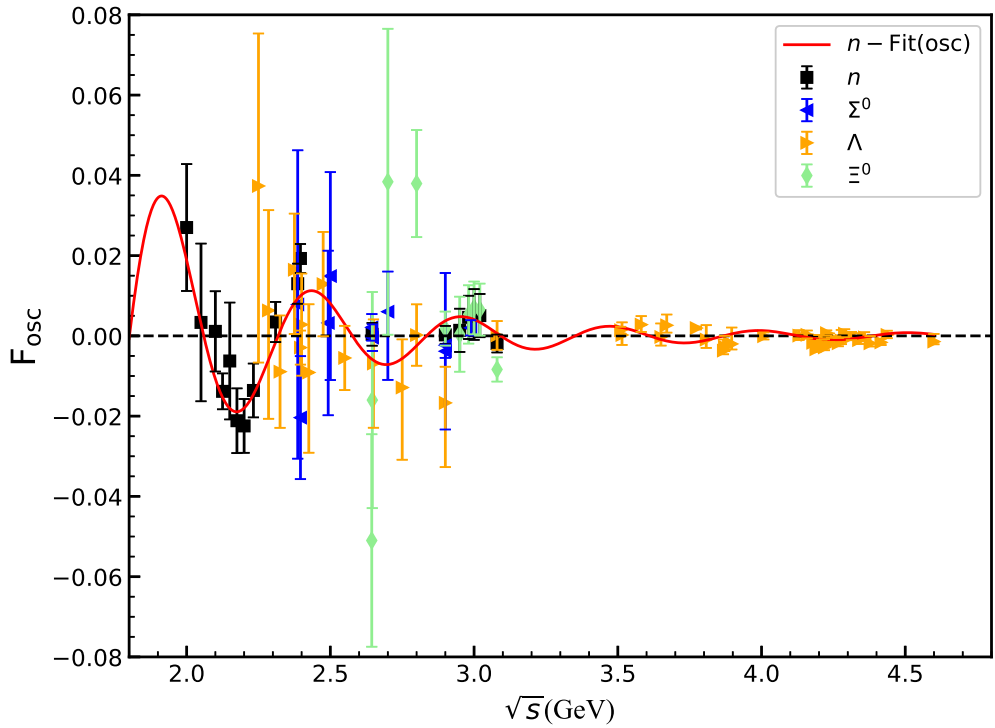}
    \caption{Oscillations of neutron, $\Lambda$, $\Sigma^0$, and $\Xi^0$ as a function of the invariant mass $\sqrt{s}$.}
    \label{fig:GoscALL}
\end{figure*}

Next, in Fig.~\ref{fig:GoscALL}, we put all the oscillation data for
neutron, $\Lambda$, $\Sigma^0$, and $\Xi^0$ together. The red-solid
line stands for the results with these fitted parameters of neutron
data. Because of large errors of the experimental data for
$\Lambda$, $\Sigma^0$, and $\Xi^0$ hyperons, one can see that the
red-line is not disagreement with these current data on $\Lambda$,
$\Sigma^0$, and $\Xi^0$ within uncertainties. This indicates that
there are unexplored intrinsic dynamics of the oscillation behavior
that need to be furthered studied.

\section{Summary}

In summary, we have considered the available experimental data on
the effective form factors of the neutral baryons neutron,
$\Lambda$, $\Sigma^0$, and $\Xi^0$ that were measured by BaBar and
BESIII collaborations. A general fit of these data are performed
with the aim to investigate the oscillation behavior of the
effective form factors. It is found that the experimental data for
neutron, $\Lambda$, $\Sigma^0$, and $\Xi^0$ are in fairly agreement
with a simple dipole function. On the other hand, our analysis does
show evidence for the oscillating features in the effective form
factors of the neutron, $\Lambda$, and $\Sigma^0$, which can be
tested by the future precise data by
BESIII~\cite{Wang:2021lfq,Mangoni:2021lmr,Xia:2021agf}. Further
improvements both on the precision of the experimental data on the
effective form factors as well as the $\sqrt{s}$ range will further
improve our understanding of the inner electromagnetic structure and
dynamics of these neutral baryons.

\section*{Acknowledgments}

We thank Profs. De-Xu Lin and Xiong-Fei Wang for useful discussions.
This work is partly supported by the National Natural Science
Foundation of China under Grants Nos. 12075288, 12135007, 11735003,
and 11961141012.

 \end{document}